# Research on Older Adults' Interaction with E-Health Interface Based on Explainable Artificial Intelligence

Xueting Huang[1*] and Zhibo Zhang[2*], Fusen Guo[3], Xianghao Wang[4], Kun Chi[5], Kexin Wu[6],

[1] Faculty of Art Design, Guangdong College of Commerce, Guangzhou, China
`milasnow0326@gmail.com`
[2] School of Engineering and Information Technology, University of New South Wales, Canberra, Australia
`z5456678@unsw.edu.au`
[3] School of Science, Computing and Engineering Technologies, Swinburne University of Technology, Melbourne, Australia
`dobbysen430@gmail.com`
[4] College of Engineering, Computing and Cybernetics, Australian National University, Canberra, Australia
`xianghao.wang@anu.edu.au`
[5] College of Professional Studies, Northeastern University, New Jersey, United States
`kaylaxchi@gmail.com`
[6] Department of Computer Science, Cornell University, New York, China United States
`kw634@cornell.edu`

**Abstract.** This paper proposed a comprehensive mixed-methods framework with varied samples of older adults, including user experience, usability assessments, and in-depth interviews with the integration of Explainable Artificial Intelligence (XAI) methods. The experience of older adults' interaction with the E-health interface is collected through interviews and transformed into operatable databases whereas XAI methods are utilized to explain the collected interview results in this research work. The results show that XAI-infused e-health interfaces could play an important role in bridging the age-related digital divide by investigating elders' preferences when interacting with E-health interfaces. Furthermore, the study identifies important design factors, such as intuitive visualization and straightforward explanations, that are critical for creating efficient Human-Computer Interaction (HCI) tools among older users. Furthermore, this study emphasizes the revolutionary potential of XAI in e-health interfaces for older users, emphasizing the importance of transparency and understandability in HCI-driven healthcare solutions. This study's findings have far-reaching implications for the design and development of user-centric e-health technologies, intending to increase the overall well-being of older adults.

*These authors contributed equally to this work and should be considered co-first authors.

# 1      Introduction

The rapid evolution of technology has profoundly impacted various aspects of society, and the realm of healthcare is no exception [1]. As the aging population in China continues to grow, coupled with the persistent challenges posed by the COVID-19 epidemic, there has been a discernible surge in the adoption of e-health solutions among older adults, as statistical evidence underscores a rise in the frequency and number of Chinese older adults engaging in e-health after the epidemic was detected [2], [3]. This phenomenon has sparked an urgent need to understand and enhance the interaction between older adults and e-health interfaces, a crucial step towards ensuring the overall well-being of this demographic.

Despite the growing adoption of e-health, older adults in China encounter substantial challenges in using these digital platforms [4]. A notable obstacle is the lack of digital literacy among this demographic, hindering their ability to recognize and effectively utilize e-health interfaces[5]. Additionally, concerns about limited accessibility and other potential usability issues further compound the difficulties faced by older adults in engaging with digital health solutions [6].

To unravel the intricacies of older adults' interaction with e-health interfaces, our research employs a comprehensive mixed-methods framework. Questionnaires will be used primarily to collect user data, specifically focusing on measuring labeling. This quantitative approach is complemented by in-depth, semi-structured interviews and observations, which are designed to capture keywords and understand the nuanced experiences, preferences, and challenges faced by older users. The real-time observations during these interviews contribute to identifying specific usability issues within the e-health interface. This multifaceted approach ensures a comprehensive exploration of older adults' engagement with e-health interfaces, combining quantitative measurements of labeling with qualitative insights derived from interviews and observations.

The collected data will all be converted into a natural language database to facilitate the application of machine learning algorithms as classifiers. The decisions made by older adults during interviews serve as the ground truth for subsequent analysis. Notably, XAI methods, such as Local Interpretable Model-agnostic Explanations (LIME) and SHapley Additive exPlanations (SHAP), are integrated to explain the decisions made by the trained classifiers. This comprehensive framework allows for a transparent analysis of the rationale behind older adults' decisions during e-health interactions.

In the culmination of this research, the outcomes will be modeled and evaluated based on the User Experience Honeycomb (except Value) framework [7]. This framework provides a structured approach to assess the holistic impact of e-health interfaces on older adults, considering factors crucial for their effective engagement and overall satisfaction.

In summary, the rise of e-health adoption among China's aging population, accelerated by the challenges of the COVID-19 epidemic, highlights the need to investigate and enhance the interaction between older adults and e-health interfaces. To alleviate dilemmas such as digital literacy and usability concerns, our research utilizes a mixed-methods approach, combining questionnaires, interviews, observations, and XAI, with outcomes assessed through the User Experience Honeycomb framework for a comprehensive understanding of older adults' engagement and satisfaction.

## 2　Literature Review

### 2.1　Older Adults and E-Health Digital Interface

The intersection of aging populations and e-health technologies is a burgeoning area of research, particularly in the context of a rapidly aging society. E-health platforms offer potential benefits in terms of accessibility and efficiency, but they also present unique challenges for older users, particularly in terms of usability and user experience [8].

The adoption of e-health services by older adults is influenced by a variety of factors. A significant challenge is the digital literacy gap. Older adults often exhibit lower levels of digital literacy compared to younger populations, which can hinder their ability to effectively use e-health services [9]. Moreover, age-related physical and cognitive declines can impact the ease with which older adults interact with digital interfaces, potentially leading to frustration and disengagement [10]. The design of e-health interfaces often fails to accommodate the specific needs of older users. Usability issues such as complex navigation, small font sizes, and inadequate instructions can exacerbate the challenges faced by this demographic [11]. Accessibility concerns, including the need for more intuitive and senior-friendly designs, are critical in ensuring that e-health technologies are equally beneficial to all age groups [12].

Traditional e-health interfaces often fall short in meeting the diverse needs of older adults, primarily due to a lack of human-computer interaction (HCI) design principles that cater to this demographic. Previous methods in HCI-based user experience surveys have typically utilized standard usability metrics, which may not fully capture the unique challenges faced by older users. These traditional methods often overlook the subtleties of how age-related factors, such as cognitive and physical impairments, affect the interaction with digital health interfaces [13]. Cham: Springer International Publishing. Moreover, conventional approaches lack a nuanced understanding of older adults' emotional responses and personal preferences when using e-health platforms, leading to a gap in truly personalized user experience design [14]. This oversight can result in interfaces that are technically sound but fail to resonate with or accommodate the practical realities of older users.

In summary, the limitations of the traditional approach to e-health interfaces and corresponding user experience surveys highlight the need for more sophisticated and resonant methods in HCI research. In response to these deficiencies, there is a need for advanced methods to explore the barriers and facilitators that influence older adults' use of e-health interfaces more accurately and transparently. We need a more inclusive, intuitive, and effective approach to user experience surveys. Such an approach would not only address the current deficiencies in the design of e-health interfaces but also inform the development of user-centered technologies that ultimately improve the well-being of the aging population.

### 2.2 Explainable Artificial Intelligence Background

The European Union's General Data Protection Regulation [15] has revealed the issue of insufficient explainability in Artificial Intelligence (AI). It emphasizes the need for understanding the rationale behind AI algorithmic decisions that adversely affect individuals. To ensure trust in the decisions made by AI systems, it is essential for AI methods to be both transparent and easily interpretable. To meet these requirements, a variety of approaches have been suggested to make the decisions of AI more comprehensible to people. These approaches, commonly referred to as "Explainable Artificial Intelligence", have been applied in numerous fields including healthcare, Natural Language Processing, and financial services [16].

Categorizations of XAI methods encompass a detailed classification of diverse techniques designed to enhance the transparency and interpretability of Artificial Intelligence, focusing on differentiating these methods based on their explanation stages, explanation scopes, and explanation output format [17]. Therefore, a more accurate and specific categorization of a single XAI technique can be achieved by viewing it from various categorization angles [18]. This approach allows for a deeper understanding and revelation of the method's features and information at multiple levels.

Whereas the main objective of HCI (Human-Computer Interaction) is to design interactions that consider and accommodate the desires and capabilities of the users [19], the utilization of XAI approaches in HCI areas could ensure that decision-makers such as developers and designers better understand the results, thereby helping users become more effective in making decisions [20]. Vicente et al. (2020) [21] conducted a study to address the scarcity of literature on how users perceive different aspects of artistic image recommendation systems, from the perspective of domain expertise, relevance, and explainability. This research explored various facets of user experience with an artistic photo recommender system from both algorithmic and HCI viewpoints. The study utilized three distinct recommender interfaces and two separate Visual Content-based Recommender (VCBR) algorithms. Zhang et al. (2023) [22] applied XAI techniques in combination with various machine learning algorithms to develop a robust explainable reputation model. This model was used for the classification of human emotion using brainwave signals. Notably, the key features influencing the assessment of emotions varied between the different XAI-based frameworks. Zhang et al. (2022) [23] employed a Convolutional Neural Network (CNN) model to categorize image spam, while post-hoc XAI methods were used to elucidate the decisions made by the black-box CNN models in detecting spam images. The findings demonstrated that the XAI-based framework achieved commendable detection effectiveness for decision-makers.

However, ensuring that E-health systems are comprehensible to their intended users remains a challenge in HCI. Adhering to user-centered design principles, it is vital to create E-health systems that are understandable and tailored to the users' specific needs and skill levels, particularly for older adults interacting with E-Health devices.

In response to these challenges, this study introduces an innovative mixed-methods approach that integrates user experience, usability evaluations, and explainable artificial intelligence (XAI) techniques. This novel approach is specifically designed to investigate how older adults interact with E-Health interfaces, addressing the unique needs of this demographic. By combining these diverse methodologies, the research aims to contribute significantly to the fields of HCI and E-Health, offering new insights into the development of more accessible and user-friendly digital health technologies.

## 3 Methodology

This paper examines the interaction of older adults with e-health interfaces, integrating principles from HCI, gerontology, and XAI. It focuses on how cognitive and physical challenges unique to the aging population affect these interactions and explores how XAI can enhance the transparency and usability of e-health systems surveys. Emphasizing user-centered design, this research aims to create accessible and intuitive e-health interfaces for older users.

The methodology of this paper adopts both quantitative and qualitative data collection approaches. Quantitative data, derived from structured questionnaires, capture a wide range of older adults' experiences with e-health interfaces, providing a substantial dataset for XAI analysis to identify trends and interaction patterns. Qualitative data, obtained from in-depth interviews and observations, offer insights into the personal and subjective experiences of older adults with these technologies. These combined methods aim to deliver a comprehensive understanding of older adults' engagement with e-health systems, informing the development of more effective and user-friendly digital health solutions.

### 3.1 Data Collection

**Survey Subjects.** From June to September 2023, this study targeted elderly residents in Guangzhou City, Guangdong Province. Employing a multi-stage sampling method, a diverse sample from two central urban and two suburban districts, encompassing 48 communities, was selected. Participants included permanent residents aged 55 and above for females, and 60 and above for males, who were in good mental health, had basic literacy skills, and consented to participate. Individuals with serious illnesses were excluded to ensure data reliability.

**Survey Instrument.** A comprehensive questionnaire, including multiple-choice and Likert scale questions, was used to assess the elderly's interaction with e-health interfaces. Likert-scale questionnaires. Open-ended questions provided insights into their personal experiences. Additionally, 40 elderly residents participated in semi-structured interviews, enriching the data with detailed accounts of their challenges and satisfaction with e-health technologies.

**Development and Validation.** The survey was tailored to the elderly lifestyle in Guangzhou, focusing on the popular WeChat Public Service app for e-health. The value of WeChat application in chronic diseases management in China. The questions, based on literature review and existing tools, were pilot-tested and refined with expert input in gerontology and HCI.
**Content of the Survey.** The questionnaire explored usage frequency, navigation ease, content readability, and overall satisfaction. Different metrics were included, aligned with the User Experience Honeycomb framework. Interview topics delved into personal barriers, difficulties encountered, and trust in digital health information.
**Administration of the Survey and Ethical Considerations.** The survey was conducted both online and in-person over a three-month period to suit the varying needs of the participants, adhering to methods suitable for diverse contexts. Clear instructions were provided to ensure understanding, with assistance readily available. Ethical standards were strictly maintained, with all participants informed about the study's purpose and their rights, including confidentiality and voluntariness. Informed consent was obtained prior to data collection, ensuring ethical integrity and adherence to research standards.

### 3.2 Explainable Artificial Intelligence Methods

As described in Section 1, the XAI methodology will provide insight and transparency into understanding older adults' interactions with the e-health interface, which will help to increase the efficiency of evaluating the user experience using the Experience Honeycomb (except Value) framework. Specifically, this work will use XAI techniques including Shapley Additive Explanation (SHAP), Local Interpretable Model-agnostic Explanations (LIME), and Anchors explainers to improve the user experience of the interaction process.

SHAP, introduced as a model-agnostic way of interpreting machine learning models, employs the concept of Shapley values. These values are derived by assessing a team's performance with and without each player, analogous to understanding each feature's impact in a machine learning model. This approach measures how the absence or presence of a feature affects the model's performance, thereby determining its influence. It helps to identify whether each feature positively or negatively influences the prediction. According to a source, SHAP values are considered a more effective method for explanation compared to traditional feature importance. Feature significance, in this context, refers to the process of assigning scores to each input attribute of a specific model, with the scores representing the importance of each feature. SHAP method conducts the explanations as the following equation:

$$g(z') = \phi_0 + \sum_{j=1}^{M} \phi_j z'_j \tag{1}$$

Where $g$ represents the model to be explained, $z' \in \{0, 1\}^M$ implies the coalition feature, $M$ means the maximized coalition size, whereas $\phi_j \in R$ stands for the attribution of specified feature j.

The main objective of LIME explainer is to develop a model that is both interpretable and locally faithful to the original classifier, while also being comprehensible to human users. In this approach, for a complex classifier, an easier-to-understand model (like a linear program) is employed, incorporating interpretable features. This simpler model is designed to closely mimic the performance of the more complex model in a local context.

The explanation model in LIME is described as $g$, belonging to a set $G$ of interpretable models that are visually presentable to users, like a linear model. The function $\pi_x(z)$ is used to measure the closeness between an instance z and x, establishing the local area around x. An objective function ξ(x) is then defined. Within this function, the L-function in ξ(x) acts as a measure of how well the interpretable model $g$ approximates the complex model f, assessed through $\pi_x(z)$ in a local context. The goal is to minimize the L-function for an optimal solution of ξ(x), ensuring that the complexity of the explanation model Ω(g) remains sufficiently low for human comprehension. And the L-function is minimized to obtain the optimal solution of the objective function when Ω(g) (the explanatory model complexity) is low enough to be understood by humans. The explanation function ξ(x) formulated by the LIME algorithm is structured as:

$$\xi(x) = \arg\ min_{g \in G}\ L(f, g, \pi_x(z)) + \Omega(g) \tag{2}$$

The method for determining the similarity degree $\pi_x(z)$ is described by the following formula:

$$\pi_x(z) = \exp\left(-\frac{D(x,z)^2}{\sigma^2}\right) \tag{3}$$

Based on the definition of the similarity degree $\pi_x(z)$ as given in equation 3, the initial objective function is reformulated in equation 4. In this equation, $f(z)$ represents the predicted outcome of the perturbed sample in the d-dimensional space (original features), considering this prediction as the response. Meanwhile, $g(z')$ signifies the predicted value in the d'-dimensional space (interpretable features). The similarity measure is then applied as a weighting feature, allowing the optimization of the aforementioned objective function through linear regression.

$$\xi(x) = \sum_{z,z' \in Z} \pi_x(z)\ (f(z) - g(z'))^2 \tag{4}$$

Anchor, a different method for interpreting classification models, was developed by Marco Tulio Ribeiro in 2018 [24]. Sharing LIME's philosophy, Anchor utilizes a perturbation-based approach to generate local explanations, capable of elucidating the specific predictions of any "black box" model. This method is characterized as being "Local", "Post hoc", and "Model agnostic". Notably, Anchor's interpretations are more aligned with human comprehension compared to other methods like LIME or SHAP, offering clear prediction rules that clarify how the model arrived at its results.

## 4      Research Process

In this section, the detailed research process of the proposed mixed-methods framework for investigating older adults' interaction with E-health interfaces is illustrated. Fig. 1 describes the overall diagram of the proposed approach. Steps including Data Collection by Questionnaires, Interviews, and Observation, Data Preprocessing and Augmentation, Data Annotation by User Experience are conducted in the previous steps to collect and process data. The processed data would be annotated from different user experience evaluation perspectives, whereas user-based classifiers would be trained based on the annotations and processed data. A user-based explainer is developed for understanding the user requirements based on the user-based classifier whereas the user experience without the explainer is also evaluated for comparison. The interview results show that user satisfaction levels improved in different user experience evaluation metrics.

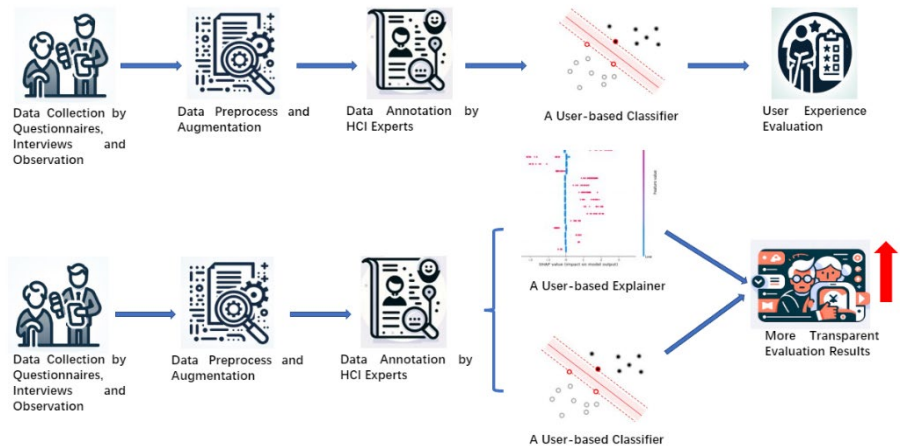

**Fig. 1.** The diagram of the proposed older adults' user experience evaluation framework.

The study conducted a comprehensive data collection exercise from June to September 2023 involving 480 elderly residents of Guangzhou City, Guangdong Province. The participant demographic comprised females aged 55 and above and males aged 60 and above, offering a representative sample for understanding the HCI and UX aspects of elderly interactions with E-health interfaces. The data showed that older users had varying levels of engagement with the e-health platform. It is notable that older users with more than a high school education were more likely to support health through e-health, which provided valuable insights into the usability and accessibility of the interface from an older user's perspective. Participants frequently cited issues such as navigation difficulty, small text size, and complex interface layouts. These findings underscore the need for an age-appropriate HCI design that accommodates the physical and cognitive changes associated with aging. Despite the challenges, a positive attitude towards the convenience and efficiency of E-health services was observed, highlighting the importance of these services in the daily lives of the elderly, especially during the pandemic.

To address potential dataset limitations and biases, and to ensure a comprehensive analysis relevant to HCI and UX research, data augmentation methods were employed. Therefore, four data augmentation methods for text data are utilized, each aimed at enhancing the diversity and volume of the collected dataset for user experience evaluation tasks. The first method, Synonym Replacement, substitutes a randomly selected word in a sentence with one of its synonyms, leveraging the WordNet lexical database for synonym-selection. The second method, Random Insertion, involves the duplication of a randomly chosen word by inserting it at another random position within the text, thereby increasing text length and introducing repetition. The third method, Random Swap, randomly interchanges the positions of two words within the text, subtly altering the sentence structure without significant impact on overall meaning. Lastly, the fourth method, Random Deletion, randomly removes words from the text with a predefined probability, simulating scenarios of missing or incomplete data. These methods collectively contribute to a more robust and varied dataset, which is crucial for training and enhancing the performance of Machine Learning classifiers and subsequent XAI explainers

The User Experience Honeycomb framework is known for its comprehensive approach to user experience evaluation. We chose the concept of this framework ( except value) to analyze the data because it is relevant to the study of user decision-making in human-computer interaction and helps to analyze in detail key areas of user experience such as Usability, Usefulness, Desirability, Findability, Accessibility, and Credibility. The following six concepts serve as a basis for analyzing the data:

**Usability** evaluates whether the system solves actual problems for the user. The paper collected data with the goal of finding out whether participants found the interface useful or not. The questionnaire asked the participant if the interface solved his/her problems. Through interviews, participants complemented the reasons why the interface solved and failed to solve their problems. In particular, the responsiveness of the interface, the level of fault tolerance, the support of multiple languages, and the availability of a specific interface channel for older adults could influence participants' decision to use or not use e-health. In addition, the layout, consistency, intuitiveness, accessibility, and availability of help channels influenced participants' decision to continue using or not using e-health.

**Usefulness** evaluates how easy the system is to use and how efficiently the user is able to complete the tasks. The data was based on the total time spent by the participants in completing the tasks using questionnaires. Additionally, during the interviews, the interviewer recorded the barriers or assistance that users encountered in interacting with the interface in order to understand which situations specifically affected the time it took for users to complete the tasks. Particularly, participants indicated that user knowledge and training had the greatest impact on the efficiency of using the e-health interface, with the role of the trainer typically being played by a junior member of the family, a peer, or a caregiver. Secondly, the performance of the device and the data also had a large impact on the efficiency of completing tasks.

**Desirability** evaluates whether the design of the interface evokes positive emotions and appreciation from the user. Data were collected using a questionnaire in which participants rated their satisfaction with each module of WeChat Public

Service's e-health (including the seven modules of medical care, health insurance, COVID-19 prevention, off-site medical care, medical information, vaccinations, and emergency centers) in order to derive the user's overall satisfaction with the interface. Usage context-based interviews and observations were used to understand which specific modules of the interface led to user satisfaction or dissatisfaction and why.

**Findability** evaluates how easy it is for users to find information and navigate the system. Data was collected using questionnaires to understand how difficult it is for participants to find specific features. Interviews and observations were used to understand which specific interaction processes and levels of interaction contribute to finding barriers and finding easiness. Physiological barriers such as poor eyesight, poor hearing, unclear speech, and shaky hands are known to contribute to the participants' finding barriers, while non-physical factors such as the participants' native Cantonese (the system is predominantly in Mandarin), the low frequency of use of e-health, and the uncertainty and fear associated with being unfamiliar with the technology mainly affect the participants' finding easiness.

**Accessibility** evaluates whether the system is easily accessible to users of all abilities and disabilities. Data were collected using interviews and observations to determine which interface design factors hinder and facilitate access to the interface for participants with different abilities (including six factors: typography and readability, contrast and color, font and icon design, navigation and menu structure, consistency of interface design, and message input methods) based on a specific context to understand which specific situations lead to low accessibility or high accessibility.

**Credibility** evaluates whether the user trusts the information and the system. Data was collected using a questionnaire to understand how much the participants trusted the system. Participants were also interviewed and observed to determine which situations resulted in low or high credibility. In particular, less exposure to digital technology, being sensitive to privacy and security issues, questioning the accuracy of health information, frustration with AI robot services, and being surrounded by fewer peers who are using e-health are factors that heavily influence participants' credibility of the interface.

## 5      Results

Using the survey data collected based on the User Experience Honeycomb framework, a Random Forest classifier is trained and then explained using XAI methods including LIME and SHAP. The visualized XAI explanation results are illustrated and shown to the interviewed older adults to evaluate the improvement of user experience from different perspectives of the User Experience Honeycomb framework. Fig. 2 to Fig. 6 show some of the survey data's explanation results of LIME explainer, Word Cloud, and SHAP explainer, whereas the detailed discussion of these XAI results and how these would help to improve user experience is described as well.

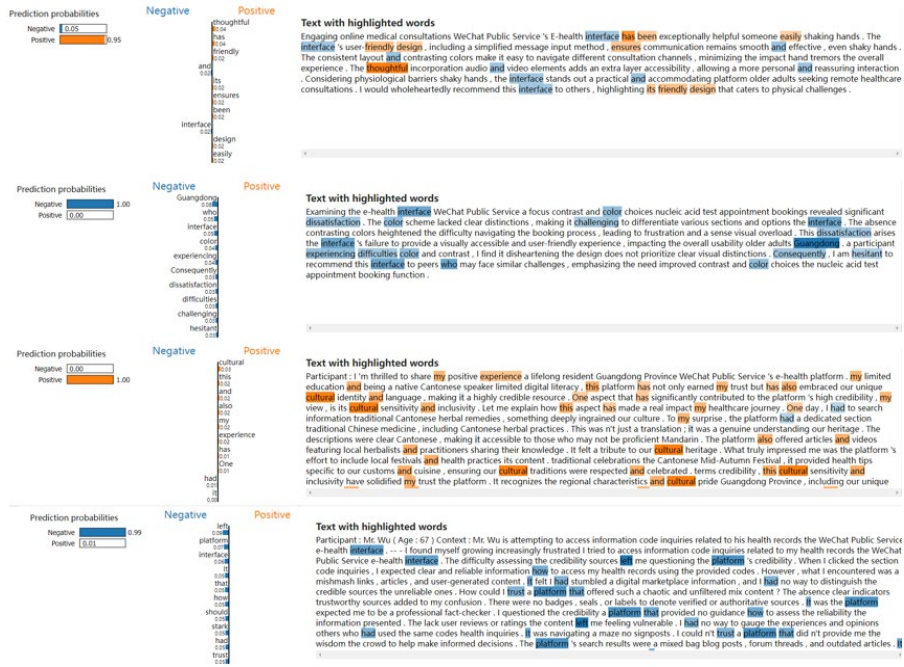

**Fig. 2.** LIME Explanation of positive and negative sample results from the user experience evaluation perspective of usability and credibility.

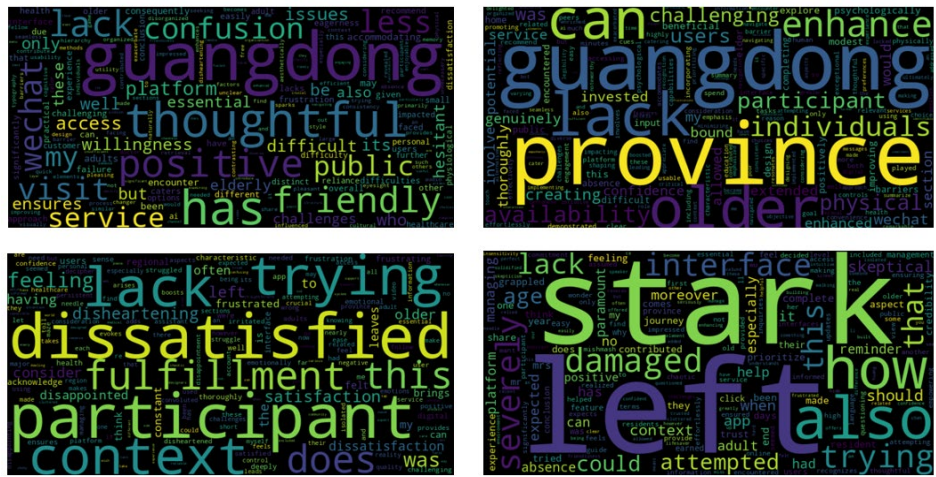

**Fig. 3.** Word Cloud Explanation results from the user experience evaluation perspective of desirability, findability, accessibility, and credibility.

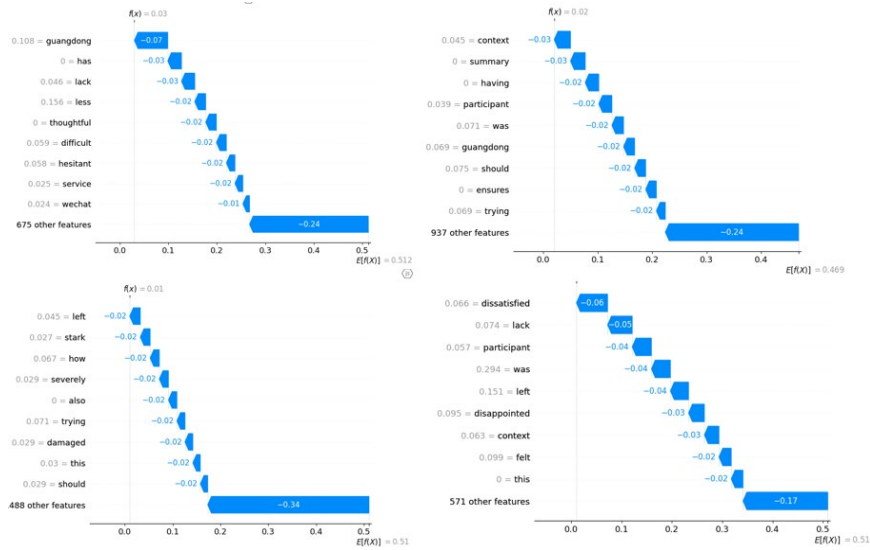

**Fig. 4.** SHAP keywords importance Explanation results from the user experience evaluation perspective of desirability, findability, accessibility, and credibility.

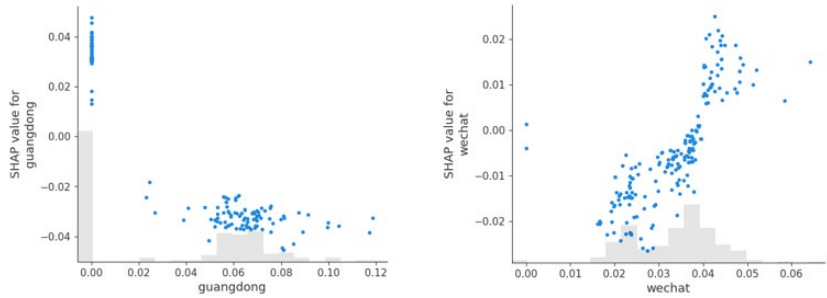

**Fig. 5.** SHAP value distribution of words "guangdong" and "wechat" from the user experience evaluation perspective of usability.

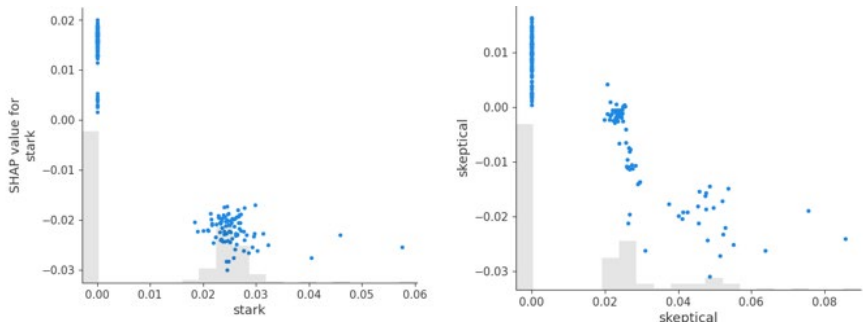

**Fig. 6.** SHAP value distribution of words "skeptical" and "stark" from the user experience evaluation perspective of credibility.

From the sample XAI results from Fig. 2 to Fig. 6, XAI explanation results could provide more transparent explanations for both designers and users. For instance, from Fig. 2, LIME Explanation of positive and negative sample results from the user experience evaluation perspective of usability and credibility, the visualized key-words that contribute to users' final decision of being satisfied or unsatisfied with usability and credibility are compared and evaluated. From this figure, it can be con-cluded that although keywords like "Guangdong", and "province" contribute a lot to the final decisions, older users tend to highlight their locations no matter whether they are satisfied or not. Apparently, when interacting with e-health, the feature that the participant's native language is Cantonese and the system's language is Mandarin received great attention, and therefore, whether the voice and text inputs support Cantonese or not became one of the determining factors of user satisfaction. On the other hand, "interface" and "platform" are keywords that affect older users negative-ly whereas "cultural" and "friendly" are keywords with positive feedback. We could conclude that much of the dissatisfaction with e-health among older participants was due to the lack of inclusive and elder-friendly interface design, and that they ex-pected culturally appropriate and user-friendly e-health products. As a result, with the help of XAI, the understanding of the user experience of older adults will improve, and both keywords related to the positive and those related to the negative will be highlighted.

From Fig. 3, the Word Cloud Explanation results from the user experience evaluation perspective of desirability, findability, accessibility, and credibility, older users, especially those with visual impairment, could understand their needs from different requirement perspectives intuitively. As for Fig. 4, it provides a significant supplement for Fig. 4 that, although some keywords play important roles in older adults' classification, users' final decision on be satisfied or not is determined more by many other factors.

From Fig. 5, the SHAP value distribution of words "guangdong" and "wechat" from the user experience evaluation perspective of usability, provide other views of user experience. While most users are not satisfied with "Guangdong" from the usa-bility perspective, not all older adult users are critical of "wechat" from the usability perspective. Although "wechat" and "guangdong" are the geographical locations and used software of all interviewed users, it can be concluded from this figure that older adult users are more dissatisfied with their geographical city rather than the popular software. On the other hand, from Fig. 6, almost all older adult users are strongly dissatisfied with the situation of "stark" from the view of credibility whereas many of them do not care about the "skeptical".

## 6      Conclusion

In advancing the field of HCI and user research, this study presents an approach to understanding the interaction of older adults with e-health interfaces, rooted in the practical application of XAI. By employing XAI techniques such as SHAP and LIME, this research provides a deeper, more nuanced understanding of how cognitive and physical changes in older adults impact their user experience with digital health platforms. This integration of XAI into user research is a significant stride, offering a more detailed and transparent analysis of user behavior and preferences. Specifically, this study highlights the necessity of language inclusivity, intuitive interface design, and cultural adaptability in e-health systems, thereby enhancing the overall user experience for older adults.

The study, however, is not without its limitations. The focus on a specific demograph-ic in Guangzhou City may limit the generalizability of the findings to other cultural and regional contexts. Additionally, the research methodology, primarily reliant on self-reporting, could be expanded in future studies to include more objective data collection techniques such as real-time user interaction tracking. This would provide a more comprehensive understanding of the user experience and help to mitigate any potential response biases. Besides, the collected data contains many meaningless features that need to be pruned in future work.

The implications of this research are substantial for the future of e-health interface design and development. By incorporating XAI into the analysis of older adults' interactions with e-health platforms, this study not only enriches decision-makers un-derstanding of user needs and preferences but also lays the groundwork for more user-centric, accessible, and empathetic digital health solutions. This approach em-phasizes the importance of tailoring e-health technologies to meet the specific needs of older users, taking into account their unique challenges and preferences. Ultimate-ly, this research contributes to a more inclusive and effective digital health environ-ment, promising to enhance the quality of life for the aging population by offering more personalized, understandable, and user-friendly e-health experiences.